\documentclass[prb,nofootinbib,twocolumn,superscriptaddress]{revtex4} 

\usepackage{graphicx}
\usepackage{dcolumn}
\usepackage{bm}
\usepackage{threeparttable}
\usepackage{times}
\usepackage{mathptmx}
\usepackage{lscape}
\usepackage{natbib}
\usepackage{amsmath}
\usepackage{braket}

\def\degree{\kern-.2em\r{}\kern-.3em}

\begin{document}


\title{ Equilibrium macroscopic structure revisited from spatial constraint }

\author{Koretaka Yuge}
\affiliation{
Department of Materials Science and Engineering,  Kyoto University, Sakyo, Kyoto 606-8501, Japan\\
}%

\begin{abstract}
In classical systems, we reexamine how macroscopic structures in equilibrium state connect with spatial constraint on 
the systems: e.g., volume and density as the constraint for liquids in rigid box, and crystal lattice as the constraint for crystalline solids. We reveal that in disordered states, equilibrium macroscopic structure, \textit{depending} on temperature and on multibody interactions in the system, is characterized by a single special microscopic structure $\textit{independent}$ of temperature and of interactions. The special microscopic structure depends only on the spatial constraint. We demonstrate the present findings providing  (i) significantly efficient and systematic prediction of macroscopic structures for possible combination of constituents in multicomponent systems, and (ii) unique and accurate determination of multibody interactions in given system from measured macroscopic structure, without performing trial-and-error simulation. 
\end{abstract}

\pacs{81.90.+c \sep 61.05.-a \sep 05.20.Gg \sep 05.10.-a \sep 02.30.Zz }

\maketitle

\textit{Introduction.}---
 Consider a classical system, where total energy ($E$) is the sum of kinetic energy ($K$) and potential energy ($U$) that is a function of  positions for constituents. 
 In equilibrium state, macroscopic structure can be uniquely specified when we define a complete set of coordinations (i.e., corresponding complete orthonormal basis functions), $\left\{q_{1},\ldots,q_{g}\right\}$. 
 Statistical physics tells us that $Q_{r}$, macroscopic structure along a chosen coordination of $q_{r}$, can be typically given by canonical average:
\begin{eqnarray}
\label{eq:can}
Q_{r}\left(T\right) = Z^{-1}\sum_{d}q_{r}^{\left(d\right)}\exp\left(-\frac{E^{\left(d\right)}}{k _{\textrm{B}}T}\right),
\end{eqnarray}
where $Z$ denotes partition function and summation is taken over possible microscopic states $d$. 
 Generally, direct determination of macroscopic structure through Eq.~(\ref{eq:can}) is nontrivial since number of possible microscopic states considered astronomically increases with increase of system size. Therefore, a variety of calculation techniques have been developed. One of the most successful techniques is Monte Carlo (MC) simulation with Metropolis algorism,\cite{metro} which samples important microscopic states mainly contributing to $Q_{r}$, and subsequent modifications have been proposed such as multihistgram method, multicanonical ensembles and entropic sampling.\cite{mc1,mc2,mc3} 
One of the approaches that does not require multiple states is coherent potential approximation,\cite{cpa} where it considers the average occupation of elements with the lack of information about geometrical structure. 
Another approach without use of multiple states is high-temperature expansion,\cite{ht} which can efficiently estimate energy as well as other physical properties at high temperature.

 When we consider a wide class of classical systems, their constituents are typically spatially constrained in various ways: For example, volume and density as the constraint for liquids in rigid box, and crystal lattice as the constraint for crystalline solids. 
Although quantitative estimations of the macroscopic structures in equilibrium state have been amply performed by a variety of 
theoretical studies,\cite{i5,i6,i7,i9,i10,i11,i13} the role of the spatial constraints on equilibrium properties does not get sufficient attention so far.  
It is thus highly expected that there remains hidden universally in equilibrium macroscopic structures, which reflect the spatial constraints. 
 Following the above considerations, we find new representation that clarifies explicit connection between equilibrium macroscopic structures and spatial constraints:
\begin{eqnarray}
\label{eq:lsi}
Q_{r}\left(T\right) \simeq \Braket{q_{r}}_{1} - \sqrt{\frac{\pi}{2}}\Braket{q_{r}}_{2} \cdot \frac{U_{r}^{\textrm{proj}}}{k_{\textrm{B}}T}.
\end{eqnarray}
Here, $\Braket{\quad}_1$ and $\Braket{\quad}_2$ respectively denotes taking arithmetic average and standard deviation over all possible microscopic states on configuration space, independently of temperature. 
$U_{r}^{\rm{proj}}$ represents potential energy of a special microscopic state, which we call ``projection state''. 
Important point here is that structure of the projection state can be determined from information of \textit{non-interacting} system.  
In other words, $\Braket{q_{r}}_1$, $\Braket{q_{r}}_2$ and structure of the projection state are determined without any information about interactions of given system or temperature, i.e., they can be known \textit{a priori} when spatial constraint is given.  


 This result reveals that in classical systems, equilibrium macroscopic structure \textit{depending} on temperature and on interactions, can be characterized by a single microscopic structure \textit{independent} of temperature and of interactions, which is a surprising fact. 
 We will see the present findings provide great advantage in practical applications: e.g., (i) efficient and systematic prediction of equilibrium macroscopic structure for possible combination of constituent elements in multicomponent systems, and (ii) unique determination of multibody interactions from measured macroscopic structure. 
 Derivation, concept, validity and applicability to practical systems are shown below.  

\textit{Derivation and concept.}---
 Before we derive Eq.~(\ref{eq:lsi}), some notations are needed. 
It is known from the equipartition theorem that canonical average for kinetic energy is determined from temperature. Therefore, to determine equilibrium macroscopic structure, only configurational average should be essential. Hereinafter, we thus focus on microscopic states on configuration space. 
 In classical systems, potential energy for any microscopic state $d$ on configuration space can be expressed by a linear combination of a complete basis functions:
\begin{eqnarray}
\label{eq:ce}
U^{\left(d\right)} = \sum_{s=1}^{g}\Braket{U|q_{s}}q_{s}^{\left(d\right)}.
\end{eqnarray}
 Here, $\Braket{a|b}$ means inner product on configuration space, corresponding to effective multibody interactions along $q _{r}$: e.g., $\Braket{a|b} = \rho ^{-1}\sum _{d}a ^{\left(d\right)}\cdot b ^{\left(d\right)}$ for discrete states and 
$\Braket{a|b} = \rho ^{-1}\int \left(a\cdot b\right) d\Omega$ ($\int d\Omega$ means integral over configuration space, and $\rho$ means normalized constant) for continuous states. 

 In multicomponent (including unary and binary) system, for a wide class of spatial constraint on the system, 
we find that density of microscopic states in terms of basis functions ($q_s$s) for almost all microscopic states can be universally given by multidimensional gaussian distribution, which leads to providing new representation of physical properties in equilibrium states.\cite{lsi} 
 Using the above facts and Eq.~(\ref{eq:ce}), we can immediately derive that density of microscopic states in terms of potential energy and a chosen coordination, $g\left(U, q _{r}\right)$, can always be given by
\begin{eqnarray}
g\left(U, q _{r}\right) \simeq \frac{1}{2\pi \left|\pmb{\Gamma}\right| ^{1/2}}\exp\left[-\frac{1}{2}\cdot\pmb{H} \pmb{\Gamma} ^{-1} \pmb{H}^{\textrm{T}}  \right], 
\end{eqnarray}
where $\pmb{\Gamma}$ denotes symmetric $2\times2$ covariance matrix for $g\left(U, q _{r}\right)$, and $\pmb{H}$ is a two-component vector of $\left(U-\Braket{U} _{1}, q _{r} - \Braket{q _{r}} _{1}\right)$. 
 We can then rewrite equilibrium macroscopic structure by using $g\left(U, q _{r}\right)$, namely, 
\begin{eqnarray}
\label{eq:taylor}
Q _{r}\left(T\right) &\simeq& Z ^{-1} \iint g\left(U, q _{r}\right) q _{r} \exp\left(-\frac{U}{k _{\textrm{B}}T}\right) dU dq _{r} \nonumber \\
&=& \Braket{q _{r}}_{1} - \frac{\Gamma _{12}}{k _{\textrm{B}}T}.
\end{eqnarray}
Since $\Braket{q _{r}}_{1}$ is constant, temperature dependence of equilibrium macroscopic structure is characterized by 
$\Gamma _{12}$, i.e., covariance between potential energy and basis function.   
 For simplicity (without lack of generality), we describe $U$ and $q _{r}$ measured from their average values, $\Braket{U}_{1}$ and $\Braket{q _{r}}_1$.  Let us take a partial average of potential energy over $g\left(U, q _{r}\right)$ only for $q _{r}\ge 0$, we get 
\begin{eqnarray}
\label{eq:pave}
U _{r}^{\textrm{proj}} &=& 2 \int _{-\infty}^{\infty} \int _{0}^{q _{r}^{\textrm{max}}} U\cdot g\left(U,q _{r}\right) dq _{r}dU \nonumber \\
&=& -\frac{2}{\sqrt{2\pi}}\Gamma _{12}\Braket{q _{r}}_{2}^{-1} \left[e ^{-\left(\frac{q _{r}^{\textrm{max}}}{\sqrt{2}\Braket{q _{r}}_{2}}\right)^{2}} -1\right] \nonumber \\
&=& \sqrt{\frac{2}{\pi}} \Gamma _{12}\Braket{q _{r}}_{2}^{-1}, 
\end{eqnarray}
where we consider a large system, $N\to \infty$ ($N$: number of constituent particles in the system). 
The last equation can be obtained since $q _{r}^{\textrm{max}}$ is the nonzero constant and $\lim _{N\to\infty} \Braket{q _{r}}_2 = 0$. 
Substituting Eq.~(\ref{eq:pave}) into Eq.~(\ref{eq:taylor}), we can vanish $\Gamma _{12}$, and obtain Eq.~(\ref{eq:lsi}). 

 Then we show that $U_{r}^{\textrm{proj}}$ corresponds to potential energy of a special microscopic state.
Let us first define that $\Braket{a}_{r}^{\left(+\right)}$ denotes a partial average of scalar quantity $a$ over all microscopic states, whose  structure satisfying $q _{r}\ge 0$.
From Eqs.~(\ref{eq:ce}) and (\ref{eq:pave}), we can rewrite $U _{r}^{\textrm{proj}}$ as
\begin{eqnarray}
\label{eq:proj}
U _{r}^{\textrm{proj}} = \Braket{\sum_{s=1}^{g}\Braket{U|q_{s}}q_{s}^{\left(d\right)}}_{r}^{\left(+\right)} = \sum_{s=1}^{g}\Braket{U|q_{s}}\Braket{q_{s}}_{r}^{\left(+\right)}.
\end{eqnarray}
The last equation can be obtained since inner product, $\Braket{U|q _{s}}$, does not by definition depend on any linear average over microscopic states on configuration space. 
Comparing Eqs.~(\ref{eq:ce}) and (\ref{eq:proj}), we can easily find that when a single microscopic state has structure of $\left\{\Braket{q_{1}}_{r}^{\left(+\right)}, \Braket{q_{2}}_{r}^{\left(+\right)},\ldots,\Braket{q_{g}}_{r}^{\left(+\right)}\right\}$, its potential energy should be identical to $U _{r}^{\textrm{proj}}$. 
We call this special microscopic state as ``projection state'', since potential energy of this state characterizes temperature dependence of equilibrium macroscopic structure projected onto a chosen coordination. From Eq.~(\ref{eq:proj}), it is clear that structure of the projection state does not depend on temperature or on interactions, since we can take partial average, $\Braket{q_{s}}_{r}^{\left(+\right)}$, without any information about temperature or energy. This directly means that when spatial constraint on the system is once given, we can \textit{a priori} know structure of the projection state.

 To derive the above equations, we only require that distribution of majority of microscopic states on configuration space for \textit{non-interacting} system is well-characterized by multidimensional gaussian. Therefore, the present finding of Eq.~(\ref{eq:lsi}) can be universally applied when this requirement is satisfied, which can be hold for wide class of classical systems. 
 We should finally make notation in derived Eq.~(\ref{eq:lsi}) for crystalline solids: Since their microscopic states are typically described in terms of static (i.e., equilibrium positions) and lattice vibrational part, slight different treatment of canonical average is required. Since the lifetime of a particular configuration is typically long enough to achieve vibrational equilibrium,\cite{ss} partition function for crystalline solids can be given in terms of the sum over microscopic states on configuration space $d$, i.e., 
$Z\simeq\sum_{d} \left[ \exp\left(-U^{\left(d\right)} - F_{\textrm{vib}}^{\left(d\right)} /k_{\textrm{B}}T\right)  \right]$ ($F_{\textrm{vib}}^{\left(d\right)}$ is vibrational free energy for configuration $d$). 
This directly means that for crystalline solids, including the effect of lattice vibration leads to just replacing potential energy $U_{r}^{\textrm{proj}}$ in Eq.~(\ref{eq:lsi}) by the sum of potential energy in equilibrium position and vibrational free energy, $F_{r}^{\textrm{proj}}$. It is now clear that structure of the projection state does not depend on whether or not vibrational effects are included: The vibrational effects can be straightforwardly included in the present theory. 

\begin{figure}[h]
\begin{center}
\includegraphics[width=0.48\linewidth]{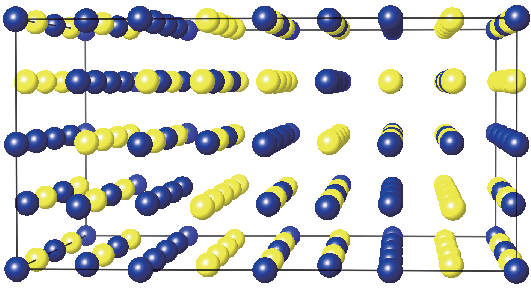}
\caption{Microscopic structure of constructed projection state along nearest-neighbor coordination on fcc lattice at equiatomic composition.}
\label{fig:1}
\end{center}
\end{figure}

\begin{figure}
\begin{center}
\includegraphics[width=0.95\linewidth]{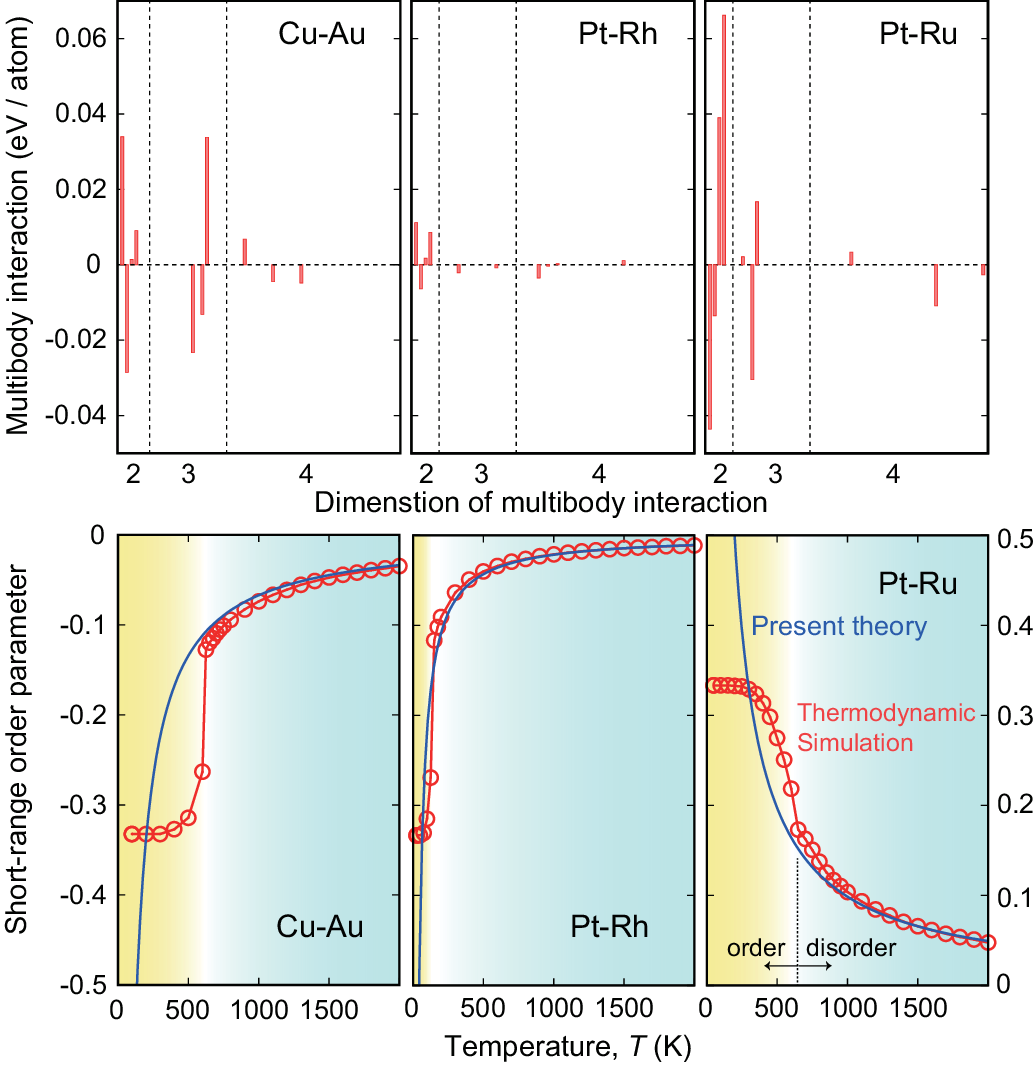}
\caption{Upper: Multibody interactions for three binary alloys of Cu-Au, Pt-Rh and Pt-Ru. Lower: Temperature dependence of equilibrium macroscopic structure along nearest-neighbor coordination, predicted by the present theory (solid curves) and full thermodynamic simulation (open circles with solid curves). }
\label{fig:2}
\end{center}
\end{figure}

\textit{Applications and discussions.}---
 We first demonstrate validity and applicability of Eq.~(\ref{eq:lsi}) to practical systems, which is compared with a full thermodynamic simulation using multibody interactions with canonical Monte Carlo simulation. We here consider substitutional crystalline solids, whose spatial constraint (i.e., crystal lattice) is much stronger than liquid or gas. 
 We employ generalized Ising model\cite{ce} providing a set of complete orthonormal basis functions to describe possible microscopic structures, where for instance occupation of a chosen lattice site $i$ is specified by Ising-like spin, $\sigma_{i}=+1$ for A and $\sigma_{i}=-1$ for B in A-B binary system.
 We choose three substitutional systems of equiatomic Cu-Au, Pt-Rh and Pt-Ru alloys, whose solid solutions are all based on fcc lattice: i.e., spatial constraints of the three systems are the same. 
 We here construct projection state along nearest-neighbor coordination. Using the Monte Carlo simulation,\cite{lsi} we obtain projection state consisting of 128-atom, where all linearly-independent pairs up to 6th nearest neighbor distance and all triplet and quartets consisting of up to 4th nearest neighbor pairs are optimized. Microscopic structure of the constructed projection state is shown in Fig.~\ref{fig:1}.
 Multibody interactions of the three alloys are obtained through first-principles calculation. Briefly, in first-principles, we calculate total energy of (i) projection state in Fig.~\ref{fig:1} for the three alloys, and (ii) to obtain multibody interactions, 302 (280 and 330) ordered structures (projection state is not included) for Cu-Au (Pt-Rh and Pt-Ru) system are applied to first-principles, then optimized twelve multibody interactions up to 4-body are extracted. Details of calculation procedures are described in our previous 
papers.\cite{lsi,pre1,pre2} The optimized interactions are shown in Fig.~\ref{fig:2}, where we find that magnitude and sign of multibody interactions are totally different for the three systems. These interactions are applied to MC simulation under canonical ensemble with 8000 MC step per site to obtain temperature dependence of macroscopic structure according to Eq.~(\ref{eq:can}). 
 Note that in Eq.~(\ref{eq:lsi}), potential energy is measured from its average, $\Braket{U}_{1}$, which is common for all coordinations. To obtain the value of $\Braket{U}_{1}$, we construct a special microscopic state whose structure have average values ($\Braket{q_{r}}_{1}$s), which is known for crystalline system as special quasirandom structure (SQS).\cite{sqs} Construction of the structure is described in detail in our previous study.\cite{lsi} Energy for SQS is also obtained through first-principles. Applying the energy of projection state measured from that of SQS to Eq.~(\ref{eq:lsi}), temperature dependence of macroscopic structure is estimated.
 The results of the present theory and of the thermodynamic simulation are summarized in the bottom of Fig.~\ref{fig:2}, where we describe structure by Warren-Cowley short-range order parameter.\cite{sro,ce-sro} It has been experimentally and/or theoretically confirmed\cite{ext1,ext2,ext5} that the three alloys take different ground-state ordered structure (yellow region in Fig.~\ref{fig:2}): L1$_{0}$ for Cu-Au, CH40 for Pt-Rh and Z2 for Pt-Ru. 
For Cu-Au alloy, reported order-disorder transition temperature ranges 650-680 K.\cite{ext5} Our thermodynamic simulation successfully predicts the reported three ordered structures as well as the transition temperature of $\sim 650$ K for Cu-Au alloy. 
We can clearly see that when the system is in disordered states (green region), the present theory successfully predict temperature dependence of the short-range order for three different alloys. 
We should emphasize here that although multibody interactions of the three alloys are totally different (upper figures in Fig.~\ref{fig:2}), temperature dependence of their equilibrium macroscopic structures are certainly characterized by a single microscopic state (projection state) that is \textit{common} for the three alloys: Validity and applicability of the present theory in Eq.~(\ref{eq:lsi}) is therefore demonstrated. 


 As seen, the present findings can avoid mapping of complex landscape of potential energy surface over possible microscopic states to determine equilibrium macroscopic structures. 
Instead, along chosen coordination, energy of merely a single special microscopic structure is required. This provides systematic and efficient prediction of macroscopic structures over possible combination of constituents, and also provides significant advantage especially when the number of components is high: e.g., the so-called ``high-entropy alloy''\cite{hea1} consisting of more than five elements at nearly equiatomic compositions. In these systems, due to the extremely high number of possible microscopic states, mapping potential energy landscape is intractable, where the present approach can overcome this problem.

\begin{figure}
\begin{center}
\includegraphics[width=1.00\linewidth]{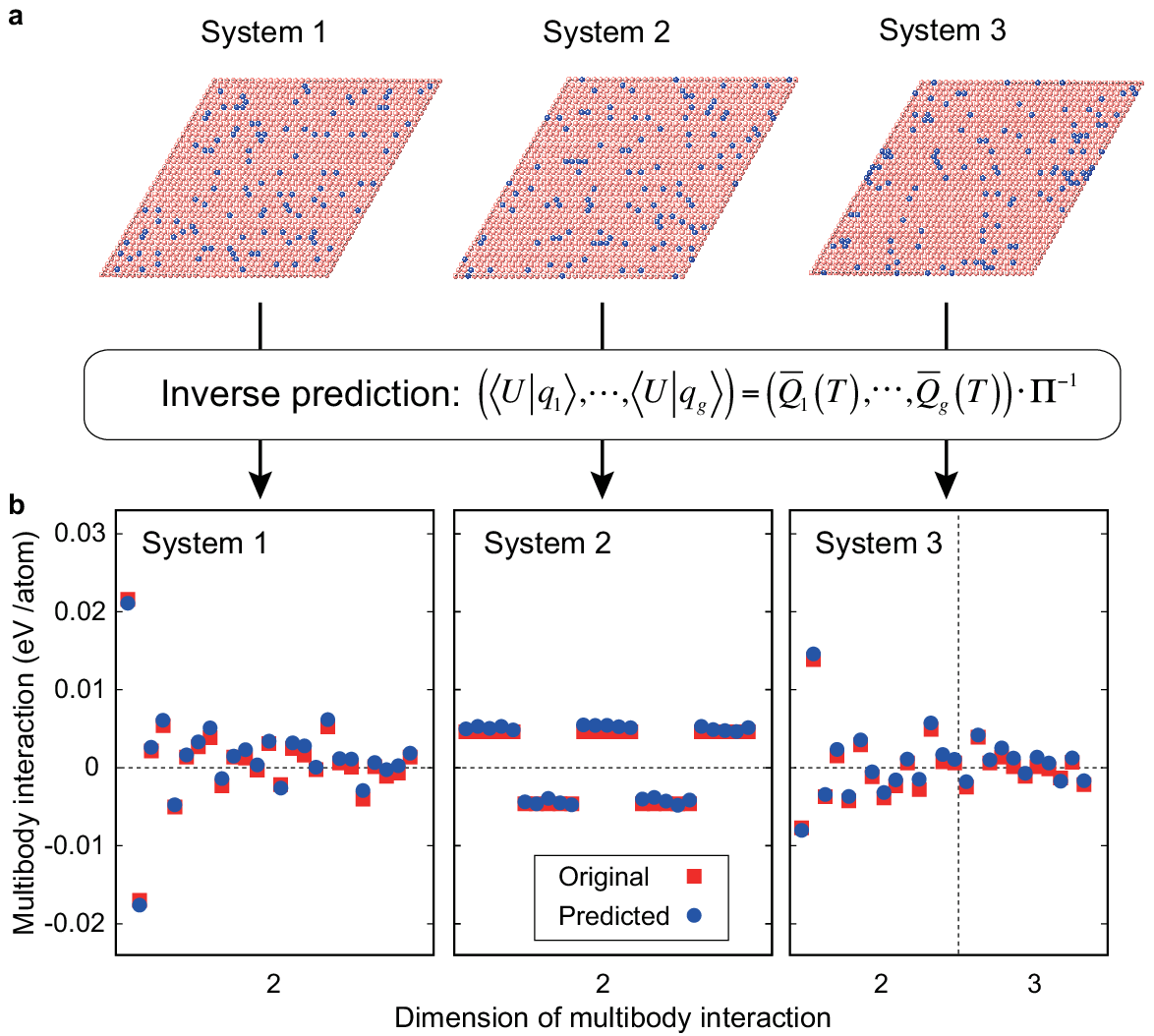}
\caption{Single snapshot of canonical MC simulation for three systems (topmost figures), using multibody interactions (closed squares in bottom figure). Predicted multibody interactions by applying the equilibrium macroscopic structures to Eqs.~(\ref{eq:inv}) and (\ref{eq:inv2}) are shown by closed circles in the bottom figure. }
\label{fig:3}
\end{center}
\end{figure}
 Next, we demonstrate another powerful application of the present theory: When equilibrium macroscopic structure at a given single temperature $T$ is once measured, then we can uniquely determine a set of multibody interactions, $\Braket{U|q_{r}}$s, in the system without performing trial-and-error simulation.

From Eqs.~(\ref{eq:lsi}) and (\ref{eq:proj}), it is easy to show the relationship between equilibrium macroscopic structure ($Q_{r}\left(T\right)$s) and multibody interactions:
\begin{eqnarray}
\label{eq:inv}
\left(\Braket{U|q_{1}},\ldots,\Braket{U|q_{g}}\right) = \left(\overline{Q}_{1}\left(T\right),\ldots,\overline{Q}_{g}\left(T\right)\right)\cdot \pmb{\Pi}^{-1}, 
\end{eqnarray}
where $\overline{Q}_{k}\left(T\right) = Q_{k}\left(T\right) - \left<q_{k}\right>_{1}$ and $\pmb{\Pi}$ is the $g\times g$ square matrix satisfying
\begin{eqnarray}
\label{eq:inv2}
\Pi_{ij} = -\frac{1}{k_{\textrm{B}}T}\cdot \sqrt{\frac{\pi}{2}}\Braket{q_{j}}_{2}\Braket{q_{i}}_{j}^{\left(+\right)}.
\end{eqnarray}
Since again, $\left<q_{k}\right>_{1}$, $\Braket{q_{j}}_{2}$ and $\Braket{q_{i}}_{j}^{\left(+\right)}$ can be known \textit{a priori} for any set of $i$ and $j$, we can construct 
matrix $\pmb{\Pi}$ without any information about multibody interactions. Therefore, multibody interactions in the given system can be uniquely determined just multiplying measured macroscopic structure by inverse of $\pmb{\Pi}$. We here note that accuracy of the predicted interactions can be easily confirmed just by applying these interactions to thermodynamic simulation to obtain equilibrium macroscopic structure. 
 To demonstrate the validity of Eqs.~(\ref{eq:inv}) and (\ref{eq:inv2}), we apply these equations to prediction of multibody interactions for binary system on two-dimensional (monolayer) triangle lattice. We artificially prepare three systems, 
whose multibody interactions are totally different: System-1 has pair interactions with gradual decrease with respect to spatial distance, system-2 is an extreme case, which has pair interactions having either single positive or negative value, and system-3 has pair and 3-body interactions. These interactions are described by closed squares in bottommost figures in Fig.~\ref{fig:3}. 
 Applying the interactions to canonical MC simulation at $T=1000$ K (i.e., full thermodynamic simulation, which is essentially the same as simulation in Fig.~\ref{fig:2}), we obtain corresponding three equilibrium macroscopic structures, where examples of single snapshot of MC simulation are shown in topmost figures in Fig.~\ref{fig:3}. 
 Then we apply these three equilibrium macroscopic structures to Eqs.~(\ref{eq:inv}) and (\ref{eq:inv2}) to predict their multibody interactions, as shown by closed circles in the bottommost figure. We can clearly see that for the three systems, multibody interactions by the present theory provide successful agreement with their original values, without performing trial-and-error simulation: Validity of 
Eqs.~(\ref{eq:inv}) and (\ref{eq:inv2}) for prediction of interactions is thus demonstrated. 


To conclude, we demonstrate that equilibrium macroscopic structures in classical systems, \textit{depending} on temperature and on multibody interactions, can be characterized by a single special microscopic structure, \textit{independent} of temperature and on interactions. 
This fact is naturally derived by clarifying how spatial constraint on the system connects with macroscopic structures in equilibrium state. 
In practical applications, we show that the present findings not only provide efficient and systematic prediction of equilibrium macroscopic structures over possible combination of constituent elements, but also enable unique determination of multibody interactions from measured macroscopic structures without trial-and-error simulation.

This work was supported by a Grant-in-Aid for Scientific Research on Innovative Areas 
``Materials Science on Synchronized LPSO Structure'' (26109710) and a Grant-in-Aid for Young Scientists B (25820323) from the MEXT of Japan, Research Grant from Hitachi Metals$\cdot$Materials Science Foundation, and Advanced Low Carbon Technology Research and Development Program of the Japan Science and Technology Agency (JST).

\end{document}